\documentclass{article}
\usepackage{spconf,amsmath,graphicx}

\usepackage{enumitem}
\setlist{nosep, leftmargin=14pt}
\usepackage{mwe} % to get dummy images
\usepackage{amsfonts}
\usepackage{multirow}
\usepackage{caption}
\usepackage{booktabs}
\usepackage{epsfig}
\usepackage{longtable}
\usepackage{rotating}
\usepackage{gensymb}
\usepackage{lscape}
\usepackage[colorlinks=true,
            linkcolor=red,
            citecolor=blue]{hyperref}
\graphicspath{{fig/}}
\usepackage{caption}
\usepackage{array, caption, floatrow, tabularx, makecell, booktabs}%
\usepackage{amsmath}

\title{Resolution- and Stimulus-agnostic Super-Resolution of Ultra-High-Field Functional MRI: Application to Visual Studies}

\name{Hongwei Bran Li$^{1}$, Matthew S. Rosen$^{1}$, Shahin Nasr$^{1}$, and Juan Eugenio Iglesias $^{1,2,3}$}
\address{\normalsize 1. Athinoula A. Martinos Center for Biomedical Imaging, Harvard Medical School, USA  \\
\normalsize 2. CSAIL, Massachusetts Institute of Technology (MIT), USA \\ \normalsize 3. Center for Medical Image Computing University College London (UCL)
London, U.K. \\
}

\begin{document}

\maketitle
% \thispagestyle{empty}
% \pagestyle{empty}
%%%%%%%%%%%%%%%%%%%%%%%%%%%%%%%%%%%%%%%%%%%%%%%%%%%%%%%%%%%%%%%%%%%%%%%%%%%%%%%%
% \vspace{-0.4cm}
\begin{abstract}
%\vspace{-0.1cm}
High-resolution fMRI provides a window into the brain's mesoscale organization. Yet, higher spatial resolution increases scan times, to compensate for the low signal and contrast-to-noise ratio. This work introduces a deep learning-based 3D super-resolution (SR) method for fMRI. By incorporating a resolution-agnostic image augmentation framework, our method adapts to varying voxel sizes \emph{without retraining}. We apply this innovative technique to localize fine-scale motion-selective sites in the early visual areas. Detection of these sites typically requires~$\le$~1mm isotropic data, whereas here, we visualize them based on lower resolution (2-3mm isotropic) fMRI data. Remarkably, the super-resolved fMRI is able to recover high-frequency detail of the interdigitated organization of these sites (relative to the color-selective sites), even with training data sourced from different subjects and experimental paradigms -- including non-visual resting-state fMRI, underscoring its robustness and versatility. Quantitative and qualitative results indicate that our method has the potential to enhance the spatial resolution of fMRI, leading to a drastic reduction in acquisition time.

\end{abstract}

%%%%%%%%%%%%%%%%%%%%%%%%%%%%%%%%%%%%%%%%%%%%%%%%%%%%%%%%%%%%%%%%%%%%%%%%%%%%%%%%
\vspace{-0.2cm}
\section{INTRODUCTION}
\vspace{-0.2cm}

High-resolution fMRI collected with ultra-high field scanners (e.g., 7T), is attracting increasing interest because of its capability to non-invasively visualize the mesoscale functional organization of the human brain. In the study of the visual cortex~\cite{kennedy2023two}, this method enables the mapping of ocular dominance columns in the striate and motion/color/stereo-selective columns in the extrastriate visual cortex~\cite{nasr2016interdigitated,yacoub2007robust,cheng2001human}. However, the improvement in spatial resolution demands increased signal averaging to counteract the decline in signal-to-noise ratio (SNR) intrinsic to smaller voxels~\cite{triantafyllou2005comparison}. Consequently, this necessitates extended scanning time, which can pose limitations in clinical and research settings due to constraints on scan time and subject comfort~\cite{vu2017tradeoffs}.

Super-resolution (SR) methods~\cite{dong2015image} that can enhance the local structure of lower resolution data acquired more quickly with reduced SNR, may offer a compelling avenue to circumvent the limitations in fMRI. Especially, deep learning-based methods~\cite{lecun2015deep} can interpolate fine details within the image~\cite{wang2020deep}, potentially reducing the necessity for prolonged scanning sessions while still achieving high-resolution fMRI data. Thus, they hold significant promise for elucidating the neural substrate with greater precision without proportionately extending the scanning time frame.

Deep learning for SR of structural brain MRI has been widely explored in the last few years~\cite{zhao2019channel,iglesias2021joint,rudie2022clinical}. One common strategy is to learn a non-linear interpolation mapping between the low-resolution and high-resolution images with deep convolutional neural networks. Typically, low-resolution images are obtained by downsampling the high-resolution ones and adding realistic artifacts (e.g., noise~\cite{wang2018esrgan}). Combined with large datasets, this strategy enables the learning of powerful image priors, parameterized by deep neural networks, by minimizing the difference between the generated SR image and the high-resolution reference image.

In the context of high-resolution fMRI, Kornprobst~\emph{et~al.}~\cite{kornprobst2003superresolution} made one of the earliest attempts to formulate fMRI SR as a convex optimization problem based on 2D local patches. Recent works~\cite{wang2020deep,ota2022super} employ 2D slices for training convolutional neural networks and validate their methods using fixed downsampling factors (which limits applicability in practice). In the 7T setting, we argue that 3D spatial information is crucial to provide stronger supervision signals than 2D slices. Furthermore, there is a lack of rigorous validation of how SR performs in cross-task and cross-subject settings, and how super-resolved images could assist fMRI analysis when considering the temporal information.

Here we propose the first method for 3D SR of 7T fMRI with deep learning to robustly improve its spatial resolution. Specifically, our approach is adaptive to varied resolutions and cross paradigms of tasks without retraining, benefiting from a domain-randomization-based data simulation strategy.

% - Single-image super-resolution offers an attractive solution to this problem, particularly when combining modern DL approaches with training data acquired with UHF MRI.

% - One paragraph of related work on SISR with DL (in the context of MRI or even more specifically brain MRI)

% - A bit of content explaining what people have tried to do with fMRI and SR (I barely found anything out there!)

% - Contributions:
% -we present a method doing blah blah capitalizing on UHF fMRI blah blah.
% -crucially, our method is agnostic to resolution to enable wide applicability. 
% - We present results on this and that task while training/testing across stimuli.
% - Also: if we're really the first ones doing this, you should totally bring it up!

\renewcommand{\thefigure}{1}
%%%%%%%%%%%%%
% \vspace{-0.1cm}
\begin{figure*}[t]
	\begin{center}
	% \vspace{0.2cm}
		\includegraphics[width=0.82\textwidth,height=0.42\textwidth]{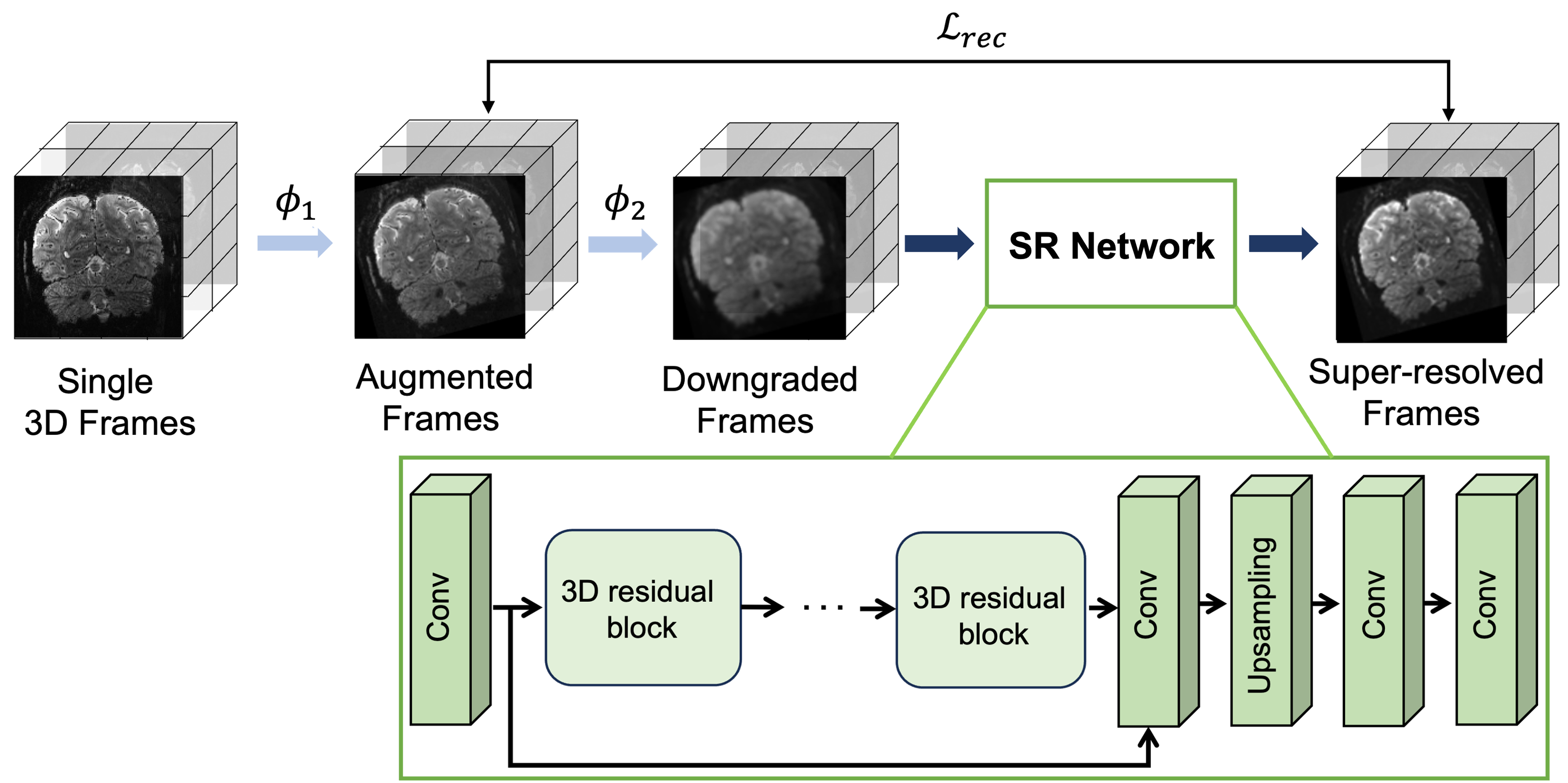}
	\end{center}
	% \vspace{-0.37cm}
    \caption{Training overview. We parse the 4D (3D+t) fMRI data to single-time-point 3D frames. We use a function $\phi_{1}$ to perform random affine and non-linear transformations, and random contrast adjustment.  We then downgrade the augmented HR frames by another function $\phi_{2}$ to perform downsampling to random lower resolutions (including a low-pass antialias filter), adding noise, and linear interpolation to the original size.}
	\label{fig:overview} 
\end{figure*}
%%%%%%%%%%%%%%%%%%%%%%%%%%%%%%%%%%%%%%%%%%%%%%%%%%%%%%%%%%%%%%%%%

% we do super-resolution with super-high-quality training data and show that it benefits downstream analyses (rather than boring direct metrics like PSNR.
\vspace{-0.3cm}
\section{Methods}
\vspace{-0.2cm}
% \emph{Loss:}
Our SR approach operates on single 3D frames to be compatible with fMRI data with varied temporal resolutions. It includes two main components: 1) domain-randomization-based data simulation, and 2) Optimization of deep neural networks using a compound loss function as shown in Fig.~\ref{fig:overview}.

\vspace{0.2cm}
\noindent\textbf{Problem definition:} \label{problem_de}
The objective is to learn a mapping from a 3D low-resolution (LR) domain (with \emph{unknown voxel size)} to a 3D high-resolution (HR) domain (with \emph{fixed} voxel size, in practice 1x1x1mm). Let $\mathcal{L}$ denote an LR image space and $\mathcal{H}$ an HR image space. In practice, $\mathcal{L}$ will have a voxel size that is common in low-resolution fMRI studies (e.g., any between 2mm and 3mm isotropic resolution). 

\vspace{0.2cm}
\noindent\textbf{Data generation:}Since obtaining both LR and HR images from the same subjects is impractical, we introduce LR domain characteristics in a generative fashion, using a model-based approach that simulates LR from HR by downsampling to a random resolution (voxel sizes uniformly sampled between 1 and 3.5mm in each direction), random rotation, brightness adjustment, and adding Gaussian noise~\cite{wang2019deep}. In summary, we aim to learn a direct resolution-agnostic interpolation function from the LR  to the HR domain $f$: $\mathcal{L} \rightarrow \mathcal{H}$.

\vspace{0.2cm}
\noindent\textbf{Loss function:} \label{loss_function}
To minimize the difference between the super-resolved image and the reference HR image, we propose a compound loss considering voxel-wise reconstruction accuracy and local structural similarity:
% \vspace{-0.1cm}
\begin{equation}
\begin{aligned}
\mathcal{L}_{rec} = \mathbb{E}_{x \sim \mathcal{L}, y \sim \mathcal{H}}[||f(x)-y||_{1} + \alpha \Delta \text{SSIM}(f(x), y)]
\end{aligned}
\end{equation}
% \vspace{-0.1cm}
where $\Delta \text{SSIM}(f(x), y) = 1 - \text{SSIM}(x, y)$. $\text{SSIM}(\cdot )$ measures the structural similarity of two images~\cite{wang2004image}. $\alpha$ is a weighting parameter of the loss terms and is empirically set to 0.2 \cite{chen2023single} in all experiments.

\vspace{0.2cm}
\noindent\textbf{Network architecture:} \label{network_architecture}
We employ a 3D fully convolutional neural network architecture inspired by~\cite{wang2018esrgan} (\emph{SR Network} in Fig.~\ref{fig:overview}). The network comprises ten dense layers connected in a residual fashion and allows for an improved flow of information and gradients with a global residual connection -- crucial for learning fine details in the SR task.

\vspace{0.15cm}
\noindent\textbf{Model inference:} \label{Inference}
%After learning the interpolation function from single 3D frames of fMRI data from one stimuli task or the resting state, we apply the trained model to 4D fMRI data acquired from different subjects under a different stimuli task. Individual frames are interpolated and concatenated in the temporal direction to yield the super-resolved 4D fMRI.
After learning the interpolation function from HR fMRI data, we apply the trained model to 4D fMRI data on a frame-by-frame basis. We then concatenate the SR frames in the temporal dimension to yield the super-resolved 4D fMRI.

\vspace{0.15cm}
\noindent\textbf{Structural and functional MRI analysis:} For each subject, inflated and flattened (i.e., structural) cortical surfaces are reconstructed based on their high-resolution anatomical data~\cite{fischl2012freesurfer}. Details of fMRI pre-processing steps, including motion correction, functional-to-structural data registration~\cite{greve2009accurate}, columnar smoothing~\cite{blazejewska2019intracortical}, and GLM modeling~\cite{penny2003mixtures} are described in Kennedy et al. (2023)~\cite{kennedy2023two}.

% For each participant, we corrected for irrelevant variances including motion parameters, global signals, and mean signals from both the ventricles and a region in the deep cerebral white matter. Following this, we derived the mean BOLD signal time course from V2/V3/V3A areas (as shown in Fig.~\ref{fig:map}), known for motion and stereo-selectivity, and used these as reference points in seed-based connectivity analysis. We then calculated the correlation coefficient between these seed time courses and the pre-processed resting-state data across all voxels in both the same and opposite brain hemispheres. 
% What statistic is the final map showing? The p value of what hypothesis?
\vspace{0.15cm}
\noindent\textbf{Analysis of regions of interest:} For each participant, regions of interest (ROIs) encompassing the visual areas V2, V3, and V3A were delineated based on the subject’s own retinotopic mapping, as detailed previously~\cite{kennedy2023two}. To enhance the sensitivity of all analyses, data from both the left and right hemispheres were combined. No hemispheric data were omitted, and all vertices within each defined ROI were incorporated into the analyses.

%
% Methods should be easy:

% -We use this architecture that has worked very well in practice as reported in the original paper.
% -We use a generator with aggressive augmentation to ensure good generalization.
% -Following our previous work We randomize the resolution to make sure it works with any resolution (here you should also explain that the idea is that you upsample whatever resolution you get to 1x1x1 and then use the network to estimate a residual)
% -We train by optimizing this loss blah blah

\renewcommand{\thefigure}{2}
%%%%%%%%%%%%%
% \vspace{-0.1cm}
\begin{figure}
	\begin{center}
	% \vspace{0.2cm}
		\includegraphics[width=1\textwidth,height=0.64\textwidth]{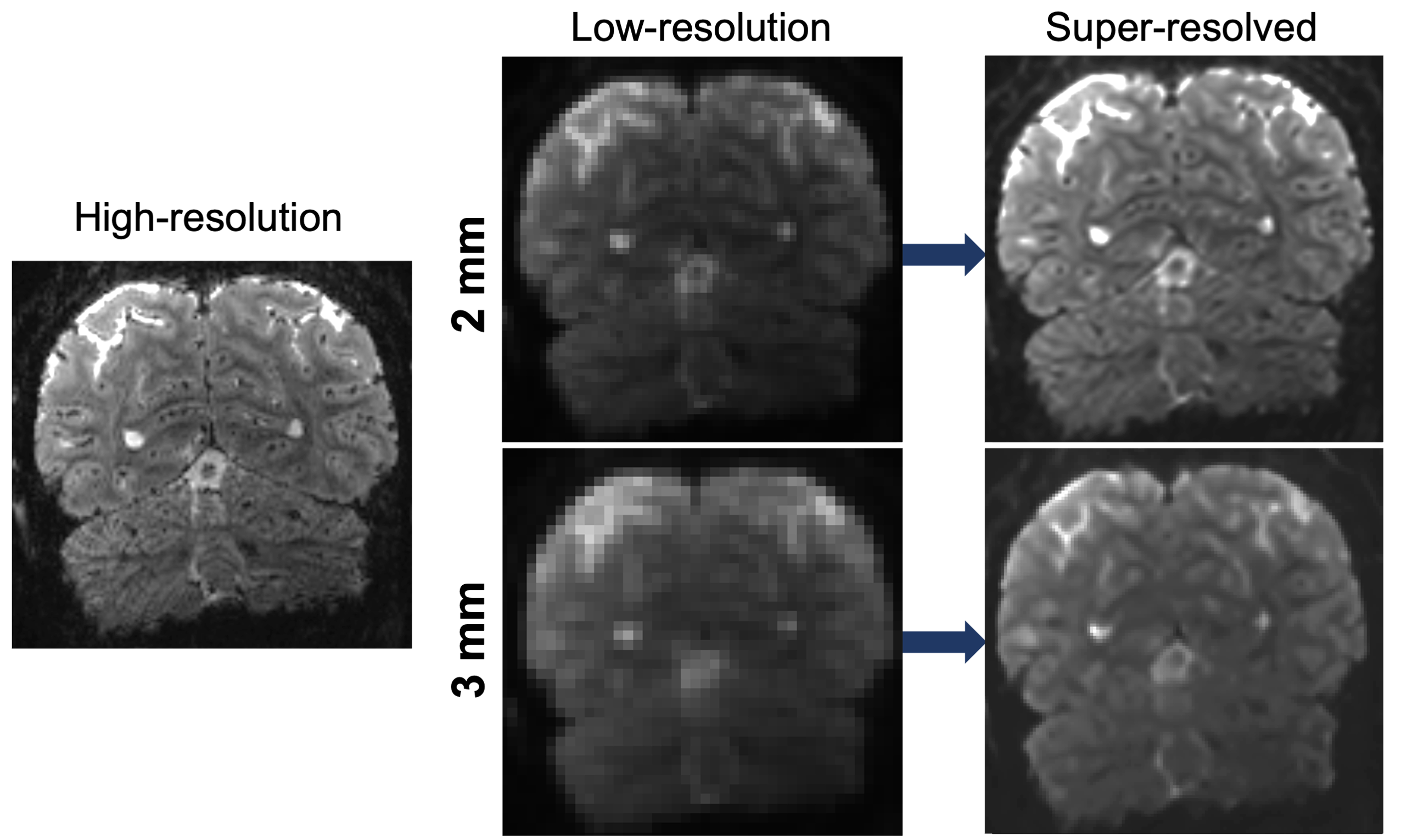}
	\end{center}
	\vspace{-0.2cm}
    \caption{Image quality enhancement by applying the SR method to low-resolution images.  Gray-white matter segregation is more apparent in super-resolved images compared to downsampled images.}
	\label{fig:sample_results} 
\end{figure}

\renewcommand{\thefigure}{3}
%%%%%%%%%%%%%
\begin{figure*}
	\begin{center}
	% \vspace{0.2cm}
\includegraphics[width=1\textwidth,height=0.46\textwidth]{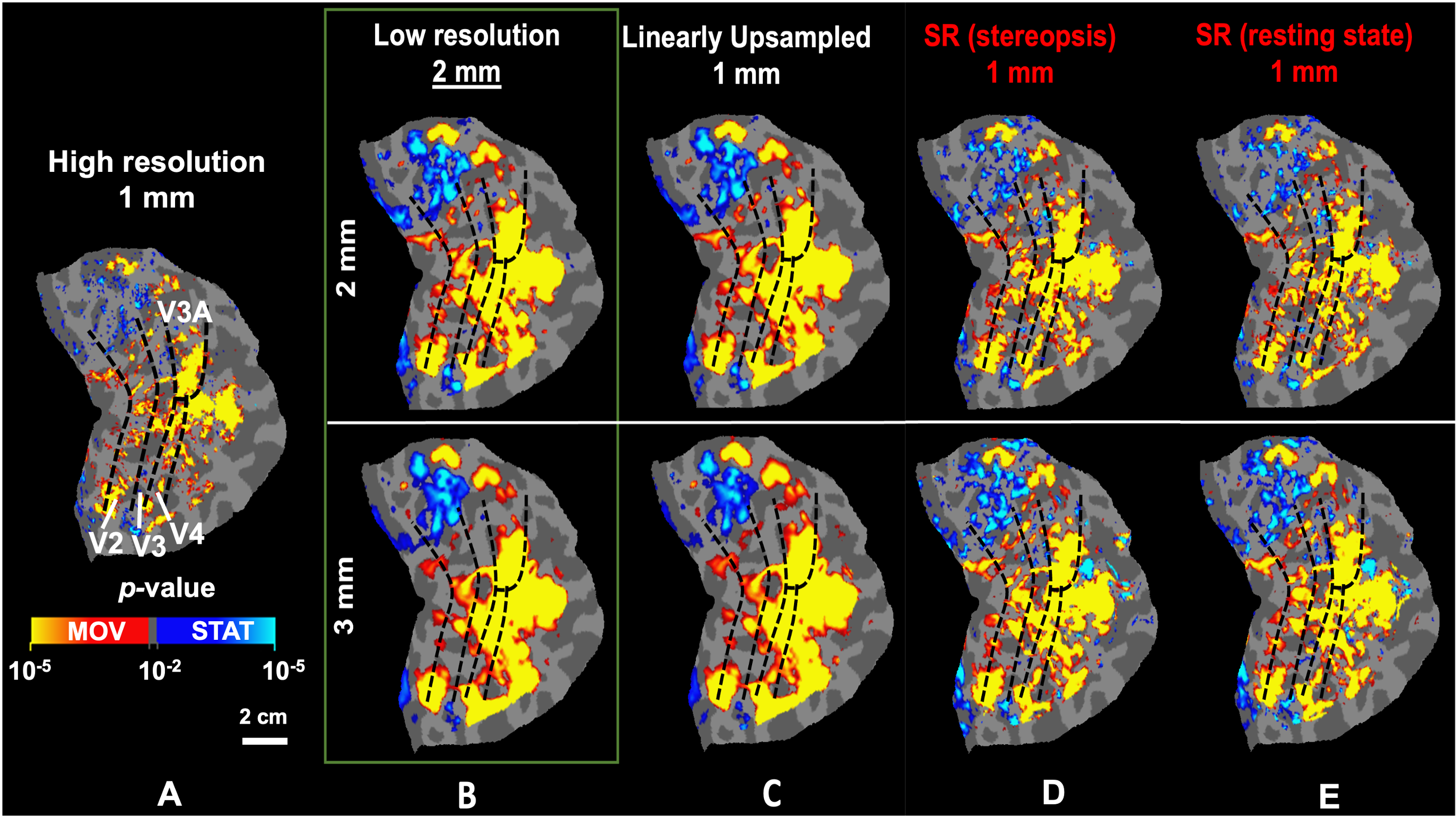}
	\end{center}
	\vspace{-0.2cm}
    	\caption{The application of the SR method improves the localization of motion-selective activity maps from low-resolution fMRI. A) Localization of motion-selective sites across V2, V3, V3A, and V4 based on the original high-resolution fMRI. B-C) Localization of the same sites based on downsampled data. Fine-scale sites are either absent or fused, causing overestimation in the size of the selective sites. D-E) Localization of motion-selective sites based on super-resolved images. The fine-scale motion-selective sites are mostly recovered in these maps. In all panels, dashed black lines indicate the borders of visual areas, defined retinotopically.}
	\label{fig:map} 
\end{figure*}
%%%%%%%%%%%%%%%%%%%%%%%%%%%%%%%%%%%%%%%%%%%%%%%%%%%%%%%%%%%%%%%%%

\section{Experiments and Results}
% \vspace{-0.2cm}
\subsection{Image Acquisition}
 \vspace{-0.1cm}
\noindent\textbf{Participants:} The study complied with NIH guidelines and had Massachusetts General Hospital IRB approval. Nine participants aged 23-44 years, all with normal or corrected-to-normal vision, standard color vision, and normal stereo vision, enrolled. They underwent multiple scans in a 7T ultra-high field scanner (Siemens Healthcare, Germany) for functional tests and were provided with written informed consent. 

\vspace{0.1cm}
\noindent\textbf{Imaging parameters:} Scans were obtained using a 7T whole-body scanner and a 32-channel head coil. Voxel dimensions were nominally 1.0 mm, isotropic. Functional images were acquired using single-shot gradient-echo EPI with the following parameters: TR=3 s, TE=28 ms, flip angle=78$^\circ$, matrix=192×192, BW=1184 Hz/pix, echo-spacing=1 ms, 7/8 phase partial Fourier, FOV=192×192 mm, 44 oblique-coronal slices, acceleration factor R=4 with GRAPPA reconstruction.
% \vspace{-0.3cm}
% \subsection{Datasets}
% \vspace{-0.2cm}

\vspace{0.1cm}
\noindent\textbf{Training set:}
In four subjects, we measured fMRI activity in separate scan sessions under (a)~resting-state conditions with eyes closed, and (b)~as subjects view random dot stereograms perceived as motion in depth vs. in the frontoparallel plane. Details on stimuli variations were in our previous work~\cite{kennedy2023two}.

\vspace{0.1cm}
\noindent\textbf{Test set:} In five subjects, we collected fMRI data as they were presented with (a) moving vs. stationary concentric rings and (b) color-varying vs. luminance-varying stimuli. These contrasts helped localize motion- and color-selective sites~\cite{goddard2020fmri}.

% \noindent\textbf{\emph{Data pre-processing}}.
% Each fMRI data obtained from each subject under one task contains 88 frames.
% All fMRI data were pre-processed and analyzed using FreeSurfer \footnote{http://surfer.nmr.mgh.harvard.edu/} and FS-FAST (v6.0)~\cite{fischl2012freesurfer}. 

% \subsection{Results}
% Qualitative results are shown in Figure 1. Figure 1 could have 2 rows: one showing a slice of a raw image before and after SR (and the ground truth); and another row showing the corresponding maps with the beautiful interdigitated stripes.
% Quantitative results are shown in Figure 2. They follow the trend of the qualitative results shown in Figure 1. 

% (if you want, you can do it the other way around: quantitative first in Figure 1, and then illustrate with a qualitative example in Figure 2)

\subsection{Experimental setup}
Two SR networks were trained on the high-resolution fMRI data acquired from stereo stimuli and resting state tasks and were applied to motion stimuli from different subjects. This allowed us to validate our approach in a cross-task, cross-subject fashion rigorously. Low-resolution images were generated by downsampling the 1mm high-resolution images to 2 and 3mm isotropic. The focus of the evaluation sought to map motion-selective sites instead of obtaining high-quality 3D frames that could be commonly quantified by quantitative metrics such as PSNR. 

\vspace{-0.2cm}
\subsection{Qualitative Results of 3D Frames}
\vspace{-0.2cm}
As Fig.~\ref{fig:sample_results} demonstrates, the gray-white matter segregation and the overall shape of sulci/gyri are notably clearer in the super-resolved images compared to the downsampled 2mm and 3mm isotropic resolution images driving the SR model. Our model effectively recovers image details across various resolutions without retraining. This enhancement is also evident in functionally mapping the motion-selective sites based on their stronger response to moving versus stationary stimuli (Fig. \ref{fig:map}A). Specifically, these fine-scale sites, mostly missing in maps based on lower spatial resolution (Fig. \ref{fig:map}B-C), are recovered in the super-resolved images (Fig. \ref{fig:map}D-E), even when using a model trained with resting-state data (\ref{fig:map}E).

\renewcommand{\thefigure}{4}
%%%%%%%%%%%%%
\begin{figure*}[!t]
	\begin{center}
\includegraphics[width=1\textwidth,height=0.42\textwidth]{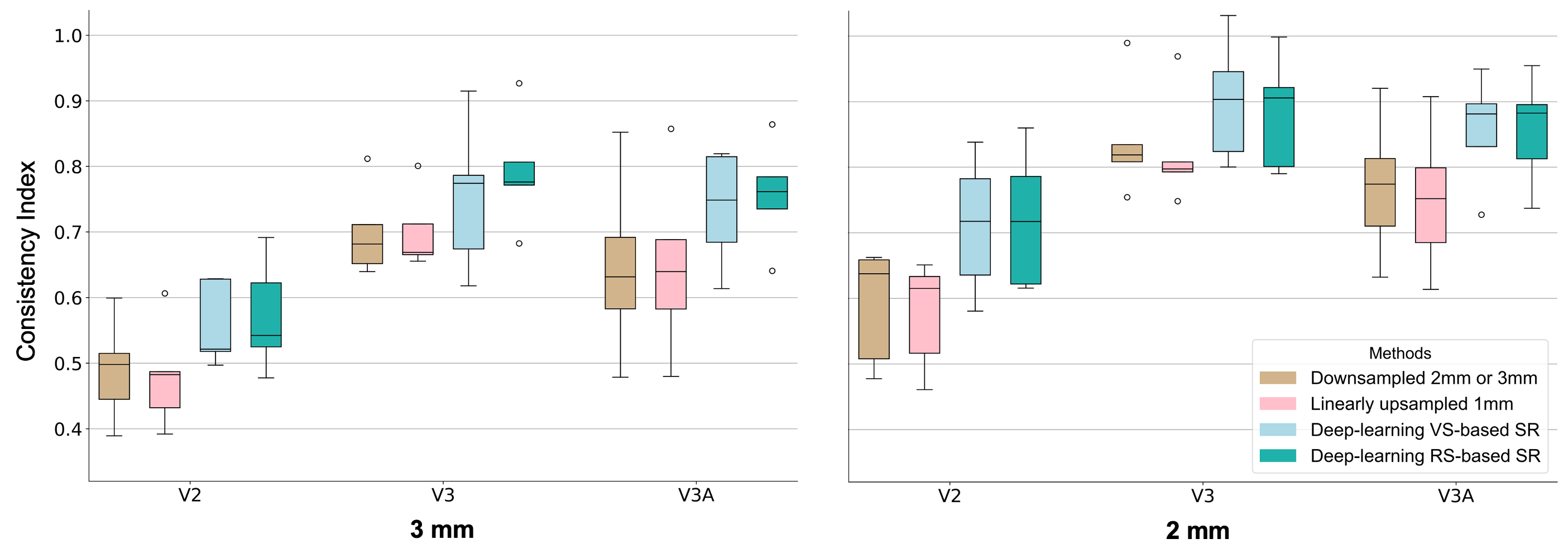}
	\end{center}
	% \vspace{-0.2cm}
    	\caption{Consistency between original motion-selectivity maps and those from downsampled, super-resolved images are shown in two panels for 2mm and 3mm resolution datasets, respectively. The ``consistency index'' represents the correlation between regenerated and original motion-selectivity maps, subtracted by the correlation with an independent high-resolution color-selectivity map. (VS=visual stimuli, RS=resting state.)}
	\label{fig:correlation} 
\end{figure*}
\vspace{-0.1cm}
\subsection{Quantitative Results of fMRI Analysis}
\vspace{-0.1cm}
To quantitatively measure the performance of the SR method, we measure the correlation between the original motion-selectivity map (Fig. \ref{fig:map}A) and the recovered maps for each ROI. These measurements are normalized relative to the correlation between the original color-selectivity map (not shown here) and the recovered motion-selectivity maps. 
%We conduct these measurements for all five subjects used in the method evaluation. 
The recovered motion-selective maps are more correlated with the original motion-selectivity than the color-selectivity map. Consistently, as shown in Fig.~\ref{fig:correlation}, the SR methods enhance the correlation between the original and recovered (compared to LR) maps across all tested ROIs. Moreover, the consistency of observations across five test subjects underscores the robustness of our SR model across subjects and stimuli.

%%The consistency between the original motion-selectivity maps, generated based on the original data, and the maps generated based on the downsampled and enhanced datasets. Panels A and B show ROI results from 2mm and 3mm resolution datasets, respectively. The consistency index represents the correlation between regenerated and original motion-selectivity maps, subtracted by the correlation with an independent high-resolution color-selectivity map. Notably, in each subject, motion- and color-selectivity maps are interdigitated relative to each other~\cite{kennedy2023two}.

%%%%%%%%%%%%%%%%%%%%%%%%%%%%%%%%%%%%%%%%%%%%%%%%%%%%%%%%%%%%%%%%%
\vspace{-0.1cm}
\section{Summary and Conclusion}
\vspace{-0.1cm}
We have demonstrated that deep learning can effectively super-resolve ultra-high-field fMRI images from lower spatial resolutions, ultimately enhancing downstream functional maps -- even across subjects and stimuli. The consistency of observations across five test subjects further highlights the robustness of our SR model. Moving forward, our research will focus on training with a larger cohort of subjects to further refine and validate this technique. Additionally, we plan to extend our exploration to include data collected inherently at lower spatial resolutions and field strengths (e.g. 3T) and compare with other image super-resolution methods such as generative adversarial works \cite{li2019diamondgan} and diffusion models \cite{song2020score}.  
% \vspace{-0.25cm}
\section{Acknowledgement}

Funded by NIH  (1RF1MH123195, R01EY030434, R01AG070-988, R01EB031114, UM1MH130981, RF1AG080371),  Jack Satter Foundation and Swiss Postdoc Mobility Fellowship. 

% \vspace{-0.2cm}
% Don't that that, please, we risk desk rejection

\bibliographystyle{ieee}
\bibliography{egbib}

\end{document}